\documentclass[twocolumn,prd,showpacs,floatfix,preprintnumbers,nofootinbib,a4paper]{revtex4}

\usepackage{epsfig}

\begin{document}

\title{Accelerated expansion and the virial theorem}

\author{Steen H. Hansen}

\affiliation{Dark Cosmology Centre, Niels Bohr Institute, University of Copenhagen,\\
Juliane Maries Vej 30, 2100 Copenhagen, Denmark}

\begin{abstract}
  When dark matter structures form and equilibrate they have to
  release a significant amount of energy in order to obey the virial
  theorem. Since dark matter is believed to be unable to radiate, this
  implies that some of the accreted dark matter particles must be
  ejected with high velocities. These ejected particles may then later
  hit other cosmological structures and deposit their momentum within
  these structures.  This induces a pressure between the cosmological
  structures which opposes the effect of gravity and may therefore
  mimic a cosmological constant. We estimate the magnitude of this
  effect and find that it may be as large as the observed accelerated
  expansion. Our estimate is accurate only within a few orders of
  magnitude. It is therefore important to make a much more careful
  calculation of this redshift dependent effect, before beginning to
  interpret the observed accelerated expansion as a time dependent
  generalization of a cosmological constant.
\end{abstract}

\maketitle

\section{Introduction}
The expansion of the universe has been observed to accelerate. This
was first noted through the analysis of supernovae \cite{perlmutter,
  riess} and more recently using cosmic microwave background and
baryonic acoustic oscillations \cite{komatsu, percival, blake}.  Most
analyses have demonstrated that a cosmological constant is in good
agreement with data \cite{larson, percival, hicken, blake2}, however,
one may easily imagine models which could allow for generalizations
beyond a simple cosmological constant. Before we start parametrizing
generalized models, or fitting new free parameters, it is important to
carefully consider and remove known effects.

One effect, which has not received much attention in this context,
arises from the merging between dark matter
structures and is always present in a bottom-up formation scenario of
cosmological structures. This effect induces an effective pressure
between cosmological structures, which in principle may lead to
systematic errors in the interpretation of the cause of the
acceleration. We will now discuss the origin of this effect.

When a dark matter structure is formed and equilibrated, either
through merging or accretion, it has to obey the virial theorem. The
virial theorem quantifies the connection between the potential energy
and the kinetic energy of the entire structure, $2K + W =0$, where $K$
is the total kinetic energy, and $W$ is the total potential
energy~\cite{clausius}.  For instance, if a structure is formed by
bringing cold particles in from infinity (i.e. initially $K_{\rm
  init}=W_{\rm init}=0$), then the virial theorem implies that an
energy corresponding to the resulting total kinetic energy, $K$, must
be disposed of \cite{binneytremaine}. For non-radiative dark matter
particles this means that a significant number of particles typically
are ejected with high momentum.

In this Letter we explain and estimate this effect. We find that it
indeed may have a significant contribution to the redshift dependent
acceleration of the expansion. Since this is an unavoidable effect, we
emphasize the need for a careful calculation of its affect on
the cosmological expansion.

\section{Effective repulsion}
Let us consider a system of 3 equilibrated dark matter structures, 2
large and 1 small. The 2 large structures are placed at a large
distance from each other, for instance at 100 times their virial radius.
The small structure, called A, is placed close to one of the large
structures, called B, for instance at 5 times the virial radius. 
A is given the circular velocity such that it is
orbiting B in a stable circular orbit. The
total force on the other large structure, called C, is given by the
total gravity from A and B.

Now, let us consider a slightly different configuration, namely one
where A is given the same speed as before, but directed towards
B. This implies that after a few dynamical times A is essentially
engulfed by B, and the 2 structures A and B will reach a new
equilibrium configuration.  The particles originating from A now sit
at a deeper potential, and they therefore have higher kinetic energy
on average.

At the end of this equilibration the virial theorem must hold, and
therefore approximately half the total change in energy must have been
disposed of. If the structures are composed of collisionless and
non-radiative particles, then that energy can only be radiated away by
``sacrificing'' some of the incoming particles, or by increasing its
size slightly. The ejected particles must leave the system with
velocities above the escape velocity. These particles will generally
leave the system in a wide cone along the axis of collision, however,
for this discussion we just assume that they leave the system
spherically.

What happens with these sacrificed particles with positive total
energy? Some will just leave the system radially, and be of no further
concern for now, however, some of them will happen to be directed
towards the other large structure, C, and when they hit C they will
transfer their momentum to it through standard gravitational effects
like dynamical friction. The effect on C is therefore an effective
pressure.

When we compare the 2 configurations above, then we see that in both
cases there will be a gravitational acceleration of C towards the
combined system of A and B. However, in the second configuration there
will be an additional pressure on C. If we knew nothing about the
merger history of A and B, then we might interpret this extra
acceleration of C as a negative gravitational effect. We will refer to
this as the {\em ``rejected acceleration''}.

If we were concerned with measuring the actual acceleration of the
expansion of the universe, then it would be important to consider the
magnitude of the effect described above, and if the effect would be
non-negligible, then a careful subtraction must be done before we can
ascribe the acceleration to e.g.\ a time varying cosmological
constant.

\section{How large is the effect}
We now wish to make a rough estimate of the effect. Let us consider
the universe today, and let us assume that half of the total mass has
been assembled into structures, all with the mass of a galaxy, $M =
10^{12} M_{\odot}$.  Naturally one should consider the full mass
distribution, $N(M,z)$, however the argument below gives the same
result for structures of $10^9 M_{\odot}$ and $10^{14} M_{\odot}$
within a factor of a few (largely from the difference in
mass-concentration), so a proper integral over $N(M,z)$ is expected to
give the same result within an order of magnitude.

Using a critical density of $\rho_c = 1.4 \times 10^{11} M_{\odot}$/Mpc$^3$,
this means that the average distance between 2 structures is about
$2.2$ Mpc.

We wish to compare the {\em rejected acceleration} 
(that is, the effective acceleration induced by the sacrificed
particles) to the acceleration which is generally believed 
to originate from a cosmological constant. The cosmological
constant induces an acceleration of the order
$GM_\Lambda(r)/r^2 \approx 9 \times 10^{-14}$ m/s$^2$ on a distance
of $r=2.2$ Mpc, where $G$ is the gravitational constant, and
$M_\Lambda(r)$ is $0.7 \, \rho_c$ integrated over a sphere
of radius $2.2$ Mpc. 
This simple estimate is, within
a factor of 2, the same as one obtains when 
numerically solving the Friedmann equation.
We therefore have that the acceleration from a cosmological constant is
given by
\begin{equation}
a_\Lambda(2.2 {\rm Mpc}) \approx 9 \times 10^{-14} \,{\rm m/s} ^2 \, .
\end{equation}

Next we consider the acceleration from the sacrificed particles.
First it is important to remember that there is a time delay from the
time when the particles were merged onto (and ejected from) one
galaxy, and till they were absorbed by another galaxy. This retarded
time depends on the typical distance between structures.  If the
ejected particles from a galaxy has of the order 200 km/s (similar to
the peak dispersion in the galaxy) then a distance of $2.2$ Mpc
implies that the particles were ejected roughly $10^{10}$ yrs ago. We
therefore see that only the particles ejected with at least 200 km/s
(in addition to the kinetic energy to leave the potential) will reach
the other galaxy today. The ejected energy may easily be even higher
or lower for some of the ejected particles. 
Numerical tests of the
ejection mechanism seems to indicate that of the order $20-30\%$
of the particles are ejected in a spherical cold collapse simulation
\cite{joyce}, and between $10\%$ and $40\%$ in merger simulations
\citep{hansenDARK}.

Let us now discuss the rate at which the particles are ejected from
the first galaxy.  The merging rate has clearly not been a constant
throughout the history of the universe, however, to get the order of
magnitude we just simplify by a linear merging rate in time, such that
$\delta M/\delta t = M/t_{\rm H} = 10^{12} M_{\odot}/13.7$ Gyr. We
therefore use $\delta M/\delta t = 10^8M_\odot /{\rm Myr}$.  To
fulfill the virial theorem, half the changed energy will be ejected
or used to increase the size of the system. 
The distribution is merger dependent, however, 
for this order of magnitude
estimate, we simply consider a quarter of the energy to  
be ejected
which implies that roughly $0.25 \, \delta M/\delta t$ will be ejected
with $v=200$ km/s. The acceleration is therefore $dv/dt = v/M \times 0.25
\, \delta M/\delta t \approx 1.8 \times 10^{-13}$ m/s$^2$.

We finally have to consider how many of the ejected particles will
deposit their momentum in another galaxy. If we consider a very long
timescale, then all particles will eventually hit some galaxy.
Equivalently, the one receiving galaxy may absorb particles from a
range of previously merging structures.  As a theoretical upper limit
we therefore have, that all the ejected particles may be absorbed.
However, many of these galaxies will lie at even larger distances, and
the particles will therefore deposit their energy far into the future.
In that case the velocities will be redshifted to lower values, and
less energy will be deposited. Typically, the velocity must be
redshifted below the escape velocity of the receiving structure
\citep{hansenDARK}.

We therefore conclude that the rejected acceleration must lie below
the value found above
\begin{equation}
a_{\rm reject}(2.2 {\rm Mpc}) < 1.8 \times 10^{-13} \, {\rm m/s^2} \, .
\end{equation}

\section{Discussions}

We have seen above that the rejected acceleration is smaller than $1.8
\times 10^{-13} \, {\rm m/s^2}$, which happens to be of the same order
of magnitude as the observed acceleration, $a_\Lambda \approx 9 \times
10^{-14} \, {\rm m/s} ^2$ within a factor of a few.

We are here entertaining the view of the redshift being a true Doppler
shift, since we are ascribing the acceleration to a changed
velocity of the galaxies and other cosmological structure.  This is
possibly completely equivalent to the view of the
expansion being a stretching of space \citep{faraoni}.  This
question is still actively debated (see a list of references in
\citep{bunnhogg}) and we will not enter that discussion here.
Similarly, it is also actively debated 
how the differences in an inhomogenneous space between a 
local accelerating space and a globally accelerating space
will manifest themselves observationally (see e.g.
\citep{clarkson,bull}).

We estimated above that the simplification from using one mass only gives
a order of magnitude uncertainty. In principle this can
be calculated more accurately, by using the correct mass distribution,
$N(M,z)$, e.g. from Press-Schechter theory. One complication that will
arise is that particles ejected from high mass structures are
difficult to absorb by smaller structures due to the higher particle
velocities. In that sense there may be a larger effect of the
rejected acceleration on larger structures.

The merging rate is both mass and redshift dependent, and therefore
the merging rate (and hence the rejected acceleration today) may be
somewhat different than the estimate used above. 

Another point which
was ignored above is that the mass of the structures were smaller at
the time when the particles were ejected. For instance the $10^{12}
M_\odot$ structure considered above would only have about have 1/4 of
that mass $10^{10}$ yrs ago. We have also been using that a quarter of
the energy must be ejected, however, if the first galaxy changes its
size more than assumed during the merging, then it may hold more
energy, and fewer particles need to be ejected, and vice versa.


A rather non-trivial point is the retarded time: the sacrificed
particles will only be absorbed in other structures at a later
time. This retardation will be different for different particles,
depending on the mass of the structure they are first merged
onto. The reason is that whereas small mass structures are formed
earlier (hence giving a larger retarded time) then the sacrificed
particles are ejected at lower velocities. This all implies that there
should be a strong redshift dependence on the rejected
acceleration. It is also interesting to note that the effect of
rejected acceleration will diminish in the far future, when structure
formation has been reduced for a long time, and all ejected particles
have either been absorbed or their energies redshifted away.

To a first approximation this effect should be isotropic, as the
observed acceleration is. However, there may be a slight anisotropic
pressure near a large over-density where most of the merging happens
along the filament. This possibility can be quantified by numerical
simulations, by observing to which degree the ejected particles are
emitted spherically or along the axes of merging.

One might ask if this effect should already have been observed by
cosmological N-body simulations, and the answer is no. Most
simulations are executed by separating the expansion of the universe
from the local effects of structure formation. This means that the
rejected acceleration cannot induce an extra acceleration of the
expansion, but at best impose an extra radial pressure on all
structures in present day simulations. This radial pressure is
low, and will be very difficult to 
disentangle from the dynamics of the normally infalling particles.
It is not known to me if a
method to perform cosmological simulations exists, where one can
dynamically include the effect of the cosmological constant: if it
will be possible to include for instance a local density dependent
acceleration due to a generalized cosmological constant, then it will
certainly also be possible to include the effect of the rejected
acceleration, at least in principle.

Finally, a different point related to the retarded time is the velocity with
which the particles are ejected from the original structures. There
will naturally be a broad distribution of velocities, since not all
particles are ejected at the same velocity. One therefore has to fold
this velocity distribution of the ejected particles with the mass
distribution, to find the full redshift dependence of this
acceleration.  One should also include the effect that different
structures emit particles at different time, which may lead to an
accumulation of the effect. For instance a $10^{10} M_\odot$ structure
will reject particles earlier and slower, than a $2 \times 10^{10}
M_\odot$ structure.  Therefore, there may be an overlap of the
arriving particles today, from the particles ejected at different
times. Furthermore, above we merely used the proper distance
today, however, in reality the relevant distance will be slightly
shorter because of the expansion of the universe.  Considering all
these effects is a somewhat non-trivial calculation which we intend to
address in the near future.

\section{Conclusion}

The observed accelerated expansion of the universe is generally
believed to be caused by a cosmological constant, or some time
dependent generalization hereof.  In this brief note we point out a
simply dynamical mechanism which may have a significant contribution
to the acceleration.

The dynamical effect arises from merging of dark matter
structures, where the virial theorem implies that a significant amount
of kinetic energy must be carried away by ejected particles. These
ejected particles will later deposit their momentum on other
cosmological structures, inducing an effective accelerated expansion.

We estimate the effective accelerated expansion from the ejected
particles to be of the same order of magnitude as the observed
accelerated expansion of the universe. The simple estimate presented
here is accurate only within a few orders of magnitude.

We emphasize that this dynamical effect is a well known physical
effect which cannot be avoided. It is therefore something which needs
to be calculated accurately and accounted for before measurements of a
time varying cosmological constant will be trustworthy.


\section*{Acknowledgement}
It is a pleasure to thank Jens Hjorth and Radek Wojtak for
discussions. The Dark Cosmology Centre is funded by the Danish
National Research Foundation.


\begin{thebibliography}{00}


\bibitem{perlmutter}
  S.~Perlmutter {\it et al.}  [Supernova Cosmology Project Collaboration],
  Astrophys.\ J.\  {\bf 517}, 565 (1999)

\bibitem{riess}
  A.~G.~Riess {\it et al.}  [Supernova Search Team Collaboration],
  Astron.\ J.\  {\bf 116}, 1009 (1998)


\bibitem{komatsu}
  E.~Komatsu {\it et al.}  [WMAP Collaboration],
  Astrophys.\ J.\ Suppl.\  {\bf 192}, 18 (2011)

\bibitem{percival}
  W.~J.~Percival {\it et al.}  [SDSS Collaboration],
  Mon.\ Not.\ Roy.\ Astron.\ Soc.\  {\bf 401}, 2148 (2010)

\bibitem{blake}
  C.~Blake {\it et al.},
  Mon.\ Not.\ Roy.\ Astron.\ Soc.\  {\bf 415}, 2876 (2011)


\bibitem{larson}
  D.~Larson {\it et al.},
  Astrophys.\ J.\ Suppl.\  {\bf 192}, 16 (2011)



\bibitem{hicken}
  M.~Hicken {\it et al.},
  Astrophys.\ J.\  {\bf 700}, 1097 (2009)

\bibitem{blake2}
  C.~Blake {\it et al.},
  arXiv:1108.2637 [astro-ph.CO].


\bibitem{clausius}
R.~J.~E.~Clausius, 
Philosophical Magazine {\bf 40} 122 (1870)


\bibitem{binneytremaine}
J.~Binney \& S.~Tremaine, 
Galactic Dynamics: Second Edition, 
Princeton university press (2008)


\bibitem{joyce}
M.~Joyce, B.~Marcos and F.~S.~Labini,
MNRAS {\bf 397}, 775 (2009)


\bibitem{hansenDARK}
S. H. Hansen et al, 2012 submitted


\bibitem{faraoni}
  V.~Faraoni,
  Gen.\ Rel.\ Grav.\  {\bf 42} (2010) 851


\bibitem{bunnhogg}
  E.~F.~Bunn and D.~W.~Hogg,
  Am.\ J.\ Phys.\  {\bf 77} (2009) 688

\bibitem{clarkson}
  C.~Clarkson, G.~Ellis, J.~Larena and O.~Umeh,
  Rept.\ Prog.\ Phys.\  {\bf 74} (2011) 112901

\bibitem{bull}
  P.~Bull and T.~Clifton,
  arXiv:1203.4479 [astro-ph.CO].



\end{thebibliography}
\end{document}